\documentclass[journal]{IEEEtran}
\newcommand{\RNum}[1]{\uppercase\expandafter{\romannumeral #1\relax}}
\usepackage{mathrsfs}
\usepackage{amssymb}
\usepackage[mathcal]{euscript}
\usepackage{diagbox}
\usepackage{cite}
\usepackage[T1]{fontenc}
\usepackage{graphicx}
\usepackage{CJKutf8}
\usepackage{makecell}
\usepackage{psfrag}
\usepackage{url}
\usepackage{stfloats}
\usepackage{amsmath}
\usepackage{array}
\usepackage{float}
\usepackage{multirow} 
\usepackage{hyperref}
\usepackage{amsthm}
\usepackage{color}
\usepackage{multicol}
\usepackage{stfloats}
\usepackage{enumerate}
\usepackage{subfigure}
\usepackage{booktabs}  

\allowdisplaybreaks[4]
\usepackage[ruled]{algorithm2e}
\usepackage{algorithmic} 
\usepackage{xurl}

\ifCLASSINFOpdf
\else
\fi
\begin{document}
\begin{CJK}{UTF8}{gbsn}
%
\title{Toward Democratized Generative AI in Next-Generation Mobile Edge Networks}



%
	\author{Ruichen Zhang, Jiayi He, Xiaofeng Luo, Dusit Niyato,~\IEEEmembership{Fellow,~IEEE}, \\ Jiawen Kang, Zehui Xiong, Yonghui Li,~\IEEEmembership{Fellow,~IEEE}, and Biplab Sikdar

\thanks{R. Zhang and D. Niyato are with the College of Computing and Data Science, Nanyang Technological University, Singapore (e-mail: ruichen.zhang@ntu.edu.sg, dniyato@ntu.edu.sg).}

\thanks{J. He, X. Luo, and J. Kang are with the School of Automation,
Guangdong University of Technology, Guangzhou 510006, China (e-mail: jiayihe@ieee.org, gdutxiaofengluo@163.com, kavinkang@gdut.edu.cn).}

\thanks{Z. Xiong is with the Pillar of Information Systems Technology and Design, Singapore University of Technology and Design, Singapore (e-mail: zehui\_xiong@sutd.edu.sg).}

\thanks{Y. Li is with the School of Electrical and
Information Engineering, University of Sydney, Sydney, NSW 2006, Australia (e-mail: yonghui.li@ sydney.edu.au).}

\thanks{B. Sikdar is with the Department of Electrical and Computer Engineering, College of Design and Engineering, National University of Singapore, Singapore (e-mail: bsikdar@nus.edu.sg).}
}
\maketitle

\begin{abstract}
The rapid development of generative AI technologies, including large language models (LLMs), has brought transformative changes to various fields. However, deploying such advanced models on mobile and edge devices remains challenging due to their high computational, memory, communication, and energy requirements. To address these challenges, we propose a model-centric framework for democratizing generative AI deployment on mobile and edge networks. First, we comprehensively review key compact model strategies, such as quantization, model pruning, and knowledge distillation, and present key performance metrics to optimize generative AI for mobile deployment. Next, we provide a focused review of mobile and edge networks, emphasizing the specific challenges and requirements of these environments. We further conduct a case study demonstrating the effectiveness of these strategies by deploying LLMs on real mobile edge devices. Experimental results highlight the practicality of democratized LLMs, with significant improvements in generalization accuracy, hallucination rate, accessibility, and resource consumption. Finally, we discuss potential research directions to further advance the deployment of generative AI in resource-constrained environments.
\end{abstract}


\section{Introduction}

The rapid growth of generative AI technologies, including large language models (LLMs) and other generative models, has led to significant breakthroughs across various domains, enabling impactful applications such as conversational systems (e.g., ChatGPT) and image generation tools (e.g., DALL-E) \cite{10679152}. These models have demonstrated remarkable capabilities in tasks such as text generation, image creation, and multimodal understanding, accelerating digital transformation in various fields such as healthcare, education, and content creation.

Despite these impressive developments, deploying large generative models remains highly challenging, particularly on mobile and edge devices. Specifically, training advanced foundation models such as GPT-4 typically requires massive computational resources, involving thousands of GPUs running for weeks, which far exceeds the capacity of conventional hardware. For example, training GPT-3, with over 175 billion parameters, requires consuming hundreds of petaflops of computational power over several weeks on advanced supercomputers\footnote{\url{https://www.trgdatacenters.com/resource/ai-chatbots-energy-usage-of-2023s-most-popular-chatbots-so-far/}}. During deployment, transferring the foundation models to mobile edge devices results in significant bandwidth consumption. During inference, running the foundation models also demands substantial computational and memory resources, surpassing the limits of most mobile CPUs and GPUs. For instance, deploying GPT-3 on a standard smartphone would be impractical due to its memory requirement of around 350 GB, whereas most mobile devices have only a few gigabytes of available memory\footnote{\url{https://www.osinto.com/post/scale-is-all-you-need}}.

Furthermore, mobile devices are subject to strict energy constraints, as executing complex AI models can rapidly drain battery life, making them unsuitable for practical use \cite{peterson2019democratizing}. For instance, running intensive AI inference tasks on a typical smartphone can deplete its battery within a few hours, which makes continuous or prolonged use unfeasible. These challenges are compounded by limited bandwidth, which makes frequent communications with cloud servers or model updates impractical in low-bandwidth environments such as rural areas. Thus, novel approaches are required to make these advanced capabilities feasible in resource-limited environments.

To address these challenges, the concept of democratized generative AI has emerged \cite{pandey2022democratized}. Democratized generative AI aims to make generative AI technologies broadly accessible by overcoming barriers such as high computational costs, energy consumption, memory limitations, and reliance on centralized infrastructure. This concept encompasses approaches such as TinyML and Small ML, which optimize models to operate efficiently on resource-constrained devices such as mobile phones and edge nodes. TinyML, for instance, emphasizes running machine learning models on small, power-efficient microcontrollers. From social and economic perspectives, democratized AI fosters inclusivity by reducing entry barriers, allowing a diverse group of users, from individual consumers to small organizations, to benefit from advanced AI capabilities without requiring extensive infrastructure\footnote{\url{https://www.splunk.com/en_us/blog/learn/democratized-generative-ai}}.

In the context of next-generation mobile and networking systems, democratized generative AI is becoming increasingly crucial. AI technologies are integral to optimizing network performance, managing bandwidth, and automating network configurations to achieve efficient communication \cite{nakao2024software}. However, deploying these resource-intensive models on mobile and edge devices, where computational power, energy, and bandwidth are often limited, presents significant challenges. Therefore, there is a need for innovative compact model strategies that make AI accessible without significantly compromising performance.

Based on these considerations, this article aims to provide a forward-looking exploration of model-centric compact model strategies for deploying generative AI on mobile and edge devices with LLMs as an example. \textit{To the best of our knowledge, this is the first article to explore a comprehensive framework for democratizing LLM deployment in mobile networks.} The main contributions of this work can be summarized as follows:

\begin{itemize}
    \item We propose a comprehensive democratized generative AI framework that incorporates compact model strategies, including \textit{quantization}, \textit{model pruning}, and \textit{knowledge distillation}, to facilitate the deployment of generative AI on resource-constrained mobile and edge devices.
    \item We conduct a systematic analysis and comparison of various compact model strategies, focusing on their ability to reduce computational demands, memory usage, and energy consumption while preserving model performance. Moreover, we define key performance metrics to assess their practical impact on generative AI applications in constrained environments.
    \item We present experimental evaluations demonstrating the effectiveness of these compact model strategies in improving generalization accuracy, hallucination rate, accessibility, and resource consumption on actual mobile edge devices. The results validate the practicality of deploying democratized LLMs on \textit{real mobile devices}.
\end{itemize}





\section{Overview of Democratized Generative AI}
\subsection{Definition}
Democratized generative AI is an effort to make generative AI technologies, such as LLMs, accessible to a broader range of users by reducing barriers such as computational costs and technical complexity. For mobile networking, the approach focuses on enabling these models to run efficiently on resource-constrained devices such as mobile phones and edge nodes. The goal is to ensure that generative AI technologies can operate without heavy reliance on centralized cloud infrastructure, allowing local devices to handle AI tasks independently and remotely. For example, consider a smartphone with a generative AI model embedded directly on the device used by a user in a rural area with limited Internet connectivity, such as a farmer analyzing a photo of a plant or an insect. The user can query an AI model such as ChatGPT directly on the device to ask about the plant’s health or identify pests for agricultural purposes. This localized processing allows users to access generative AI or multimodal LLM features such as image generation or enhancement in real time, regardless of network conditions\footnote{\url{https://www.nvidia.com/en-us/autonomous-machines/embedded-systems/}}. Generally, the benefits of democratized generative AI are as follows:
\begin{itemize}
     \item \textbf{Increased Accessibility}: Democratized generative AI significantly improves accessibility by making advanced AI tools available to a broader range of users, including those in remote or resource-constrained environments. By enabling local processing on mobile devices, more users, especially in rural areas, can access generative AI features without relying on robust cloud infrastructure or high-bandwidth Internet connections. This approach democratizes AI, allowing individuals such as farmers, students, or small business owners to leverage powerful AI tools in daily tasks without depending on centralized cloud systems\footnote{\url{https://ai.google.dev/edge/litert}}.

    \item \textbf{Improved Privacy}: Democratized generative AI enhances privacy by processing sensitive data locally on the device, preventing the need to transmit personal information to cloud servers. For example, Apple's "on-device intelligence" for features including Face ID and Siri processes facial recognition and voice commands directly on the device, significantly reducing the risk of data breaches and enhancing user trust\footnote{\url{https://www.apple.com/privacy/}}.
    
    \item \textbf{Offline Functionality}: Democratized generative AI enables users to access AI capabilities even in areas with poor or no Internet by processing tasks locally. For instance, Google Translate's offline mode allows users to translate text and conversations without an Internet connection, thanks to on-device AI models\footnote{\url{https://support.google.com/translate/answer/6142473?hl=en}}. This ensures that users can continue using AI tools regardless of network availability.
\end{itemize}

\begin{figure*}[t]
\centering{\includegraphics[width=0.88\textwidth]{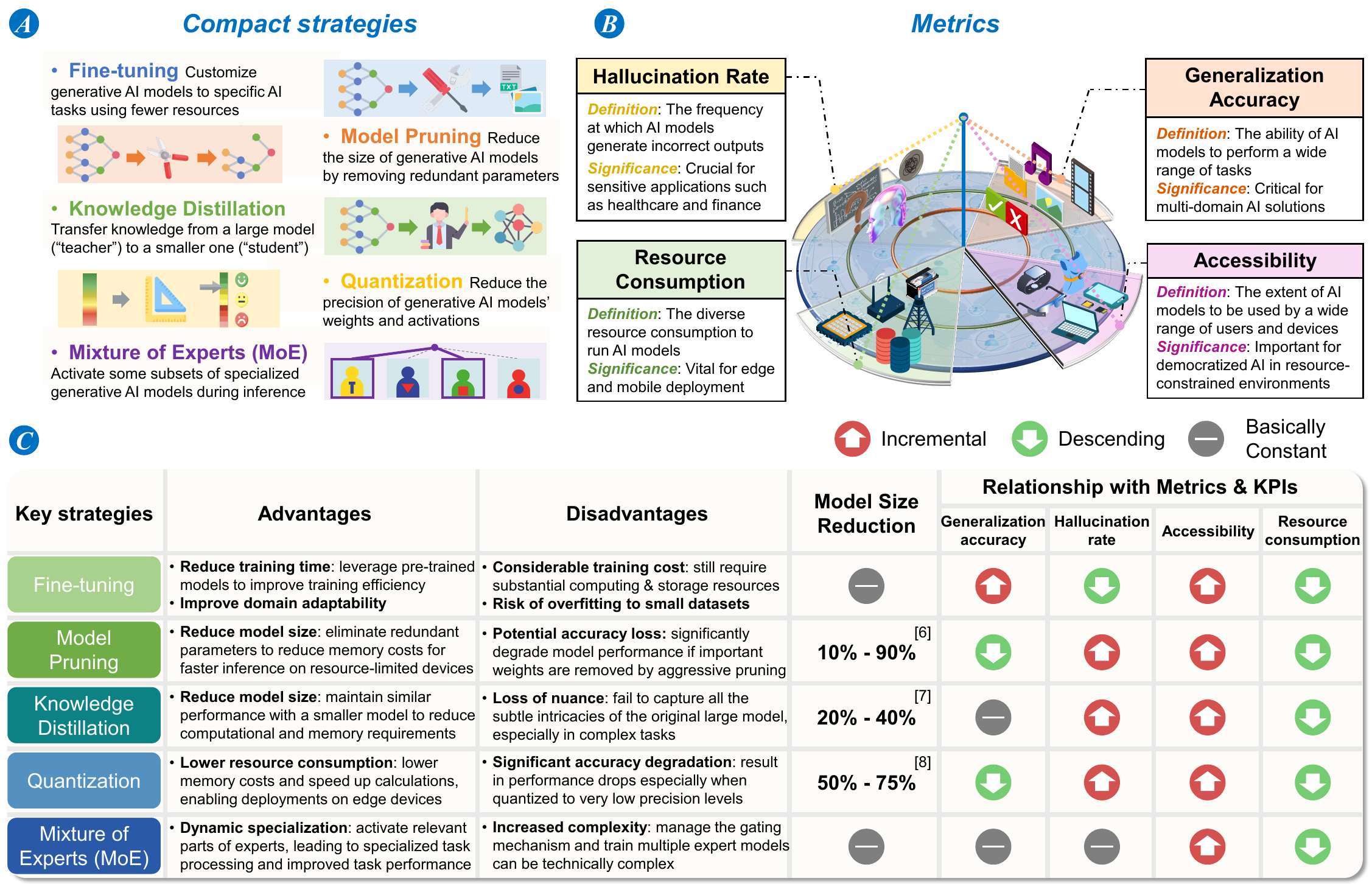}}
\caption{Overview of democratized generative AI. (A) Key compact model strategies, including fine-tuning, model pruning, distillation, quantization, mixture of experts, and caching. (B) 4-dimensional evaluation for democratized generative AI highlighting metrics including energy efficiency, hallucination rate, generalization accuracy, and accessibility. (C) Table showing the impact of each strategy on the metrics, with arrows indicating whether the effect is incremental, descending, or basically constant. }\label{Fig1}
\end{figure*}

For a more intuitive understanding, we present a comprehensive overview of democratized generative AI, which is depicted in Fig. \ref{Fig1}.

\subsection{Compact Strategies for Democratized Generative AI}

Several compact model strategies have been developed to deploy democratized generative AI models in resource-constrained environments. These strategies aim to reduce AI models' computational and memory footprints while maintaining high performance on mobile and edge devices.

\subsubsection{Fine-Tuning}

Fine-tuning involves adapting a pre-trained AI model to specific tasks or environments, reducing the need for training from scratch. In mobile and edge environments, fine-tuning models such as GPT or BERT allows them to handle specific tasks such as bandwidth management and real-time network optimization. By refining the pre-trained model for these specific applications, fine-tuning ensures the model remains flexible while improving efficiency for dynamic operational settings. Typically, fine-tuning does not inherently reduce the model size since it retrains the entire model for target tasks while keeping the original architecture intact. However, advanced fine-tuning techniques such as parameter-efficient fine-tuning (PEFT) can be used to adjust only a subset of the model's parameters, significantly reducing the number of trainable parameters without modifying the full model. This approach includes methods such as Adapters, Low-Rank Adaptation (LoRA), and Prefix Tuning, which allow fine-tuning to focus on more important parts of the model while maintaining overall performance. For example, \cite{devlin2019bert} demonstrated that fine-tuning BERT for specific tasks, such as language translation, reduced the computational demand by approximately 50\% compared to training from scratch. Furthermore, using PEFT techniques could reduce the number of trainable parameters by up to 90\%, making the model more lightweight and adaptable for resource-constrained devices.

\subsubsection{Model Pruning}
Model pruning eliminates redundant or unimportant parameters, reducing the model's size and computational complexity with minimal impact on performance.  In mobile and edge environments, pruning is especially effective for reducing the memory footprint of LLMs, allowing them to run on devices with limited processing power. By trimming down less critical parts of the model, pruning ensures that real-time inference can still be achieved without sacrificing significant accuracy.  For example, \cite{blalock2020state} showed that pruning neural networks resulted in up to 90\% model size reduction while maintaining almost identical performance, making this strategy particularly advantageous for real-time generative AI applications on mobile devices.


\subsubsection{Distillation}

Distillation involves training a smaller, more efficient ``student" model to replicate the behavior of a larger ``teacher" model. In mobile and edge environments, distilling large models such as GPT into smaller versions allows these models to efficiently handle tasks, e.g., natural language understanding or network log analysis, without relying on cloud-based services. This strategy not only reduces the size of the model but also maintains the performance required for high-quality decision-making in constrained environments. For example, \cite{sanh2019distilbert} demonstrated that by distilling BERT into a smaller model (DistilBERT), the model's size and computational load were reduced by 40\%, while retaining 97\% of BERT’s language understanding capabilities.

\subsubsection{Quantization}

Quantization reduces the precision of model weights and activations, thereby decreasing memory usage and computational costs. This compact model strategy is particularly beneficial in mobile devices with limited computational resources or without GPU. By converting 32-bit floating-point operations to 8-bit integers, quantization significantly improves both speed and energy efficiency without a substantial loss in accuracy. For example, \cite{jacob2018quantization} illustrated how quantizing deep learning models reduced memory usage by up to 75\%, making models such as GPT and BERT more suitable for real-time applications in mobile environments where energy consumption was critical.

\subsubsection{Mixture of Experts (MoE)}

Mixture of experts (i.e., MoE) is an architectural design that activates only specific parts of a model (i.e., experts) based on the input, reducing the overall computational burden. In mobile and edge environments, MoE allows models to utilize resources efficiently by using only the necessary components for a given task, improving both processing speed and power consumption. For example, \cite{shazeer2017outrageously} highlighted that MoE reduced the computational budget by up to 90\% by activating only a subset of the network’s experts for each task.

\subsection{Performance Metrics}
Evaluating the performance of democratized generative AI models in mobile and edge environments requires more than accuracy. Several other metrics are critical for understanding their effectiveness in these constrained environments.

\subsubsection{Generalization Accuracy}
Generalization accuracy refers to the ability of an AI model to perform well across various tasks and scenarios. In mobile and edge environments, maintaining generalization accuracy is particularly important, as models are often downsized to fit the resource constraints of a target device. While compact model strategies, including pruning, distillation, and quantization, can reduce the size and complexity of models, they must also ensure that the models retain the ability to generalize across different tasks without a significant drop in performance. For example, \cite{sreenivas2024llmpruningdistillationpractice} highlighted that although pruning and distillation reduced model size, it preserved most of the original model’s performance. In fact, A model with only 4B parameters obtained through pruning and distillation of the Llama-3.1-8B model retained 95\% of accuracy performance while reducing the model size by 47.5\%, demonstrating that it is possible to achieve a good balance between size reduction and generalization accuracy. 

\subsubsection{Hallucination Rate}
Hallucination rate refers to the frequency at which AI models generate incorrect or nonsensical outputs. This metric becomes particularly crucial as models are downsized for deployment on mobile and edge devices. In mobile and edge environments, a reduced model may experience a higher hallucination rate due to excessive reduction in model size and complexity. Evaluating this metric is essential to ensure that downsized models still produce reliable and accurate outputs. For example, SelfCheckGPT~\cite{manakul2023selfcheckgpt} provided a framework for evaluating hallucination in generative models. It used techniques including BERTScore and consistency checking between stochastically generated outputs to assess whether the generative model’s responses were factually correct. In mobile and edge environments, this approach could be adapted to assess the hallucination rate of downsized models, especially without access to extensive models on the cloud.

\subsubsection{Accessibility}

Accessibility refers to the extent to which AI models can be deployed and used by a wide range of users, particularly in resource-constrained environments. In mobile and edge environments, accessibility is enhanced when models require fewer computational resources and are less dependent on robust infrastructure. Democratized AI must ensure that models are accessible to diverse users who lack extensive cloud or server capabilities. For example, \cite{7460664} discussed how combining compact model strategies, such as quantization and pruning, enabled AI models to run efficiently on resource-constrained devices. These strategies helped reduce memory requirements by up to 57\% with less than 1\% performance loss, making AI more accessible to a wide range of users, including small businesses and developers without access to high-end hardware.

\subsubsection{Resource Consumption} 
Resource consumption encompasses multiple factors critical for mobile and edge environments, including energy efficiency, computational load, communication overhead, and memory usage. In addition, to reduce other resources' consumption, minimizing communication overhead is crucial in mobile environments, where models must process data locally to reduce the frequency of large data transfers to cloud servers. For example, \cite{laskaridismobile} showed that quantization techniques enabled a phone to process up to 590 prompts before it ran out of power while also lowering memory and computational demands, making AI models more suitable for mobile devices where both power and resource efficiency are critical.

Additional strategies and metrics are needed to be considered that cater to specialized, localized applications. For instance, caching can be leveraged to improve response times and efficiency, especially in constrained or resource-limited environments. Additionally, metrics focused on personalized and context-specific performance such as responsiveness in highly customized applications are essential. Examples such as LLMs tailored for agricultural queries or designed as travel guides within vehicular networks demonstrate the value of AI models that meet distinct needs in a democratized and accessible way, ensuring usability and relevance across diverse user bases.
\begin{table*}[t]
    \centering
    \caption{Comparison of Key Compact Model Strategies.}
    \label{table}	
    \includegraphics[width=0.95\textwidth]{Systemsmodel/table1_v3.pdf}
\end{table*}

\section{Comparative Analysis of Key Compact Generative AI Strategies}

In this section, we comparatively analyze the key strategies for democratized generative AI. We first compare the compact model strategies from multiple perspectives, including training methods, applicable network types, user demands, and generative AI applications. We also present practical application examples based on these strategies, as shown in Table \ref{table}. Then, we discuss the representative integrative networking under the synergy of different compact model strategies.

\subsection{Comparison of Key Compact Model Strategies}
\subsubsection{Fine-tuning} Fine-tuning focuses on task-specific adaptations rather than model simplification. In edge networks, PEFTs, such as LoRA or Prefix Tuning, are particularly useful for keeping resource consumption low, making it suitable for personalized applications such as healthcare diagnostics or AI-driven virtual assistants. Compared with pruning and quantization, fine-tuning does not achieve model lightweight directly but rather democratizes generative AI by enabling models to address specific domain tasks. For example, GitHub Copilot\footnote{\url{https://github.com/features/preview/copilot-customization}} is now available for users to fine-tune the model and make it a domain-specific personal AI assistant.

\subsubsection{Pruning} Pruning focuses on simplifying the structure of the model without changing the model weights~\cite{huang2024large}. In edge networks, pruning removes redundant neurons from the model, allowing it to perform tasks on resource-constrained devices efficiently. Pruning differs from quantization in that it reduces the number of active components rather than lowering the precision of computations. For example, LLM-Pruner\footnote{\url{https://github.com/horseee/LLM-Pruner}} realizes LLaMA/Llama-2 structural pruning with 20\% parameters pruned, making LLMs more available on edge devices.

\subsubsection{Knowledge Distillation} Distillation focuses on transferring knowledge from large models to smaller ones. It is commonly used after quantization and pruning to reduce the performance loss caused by these lightweight methods~\cite{zou2024genainet}. Compared with pruning and quantization, knowledge distillation achieves a better balance between performance and efficiency. For example, the mobileBERT\footnote{\url{https://github.com/google-research/google-research/tree/master/mobilebert}} model distills knowledge from a teacher model, i.e., BERT\textsubscript{LARGE} model, making it suitable for efficient deployment and inference on lightweight devices.

\subsubsection{Quantization} Quantization focuses on reducing memory footprint and improving computational efficiency. Quantization is beneficial at the physical layer, particularly for real-time applications where ultra-low latency and energy efficiency are prioritized, such as real-time Question-and-Answer (Q\&A) or AR/VR applications. Unlike pruning, quantization treats all parameters equally without considering the importance of each parameter, thus drastically reducing memory and computational requirements. This makes it essential for mobile and IoT devices operating under stringent power constraints. For instance, AWQ\footnote{\url{https://github.com/mit-han-lab/llm-awq}} compresses multiple LLMs for fast inference, enabling generative AI applications on mobile devices in real time.

\subsubsection{MoE} MoE focuses on task distribution across the system rather than reducing the internal model complexity. MoE is particularly suitable for applications in collaborative environments where diverse tasks are distributed across multiple devices. MoE differs from other methods in its great flexibility and scalability. For example, Uni-MoE\footnote{\url{https://github.com/HITsz-TMG/UMOE-Scaling-Unified-Multimodal-LLMs}} adopts an MoE architecture to process multimodal tasks, with each expert specializing in a specific modality, making it well-suited for deployment in collaborative networks.

\begin{figure*}[t]
\centering{\includegraphics[width=0.98\textwidth]{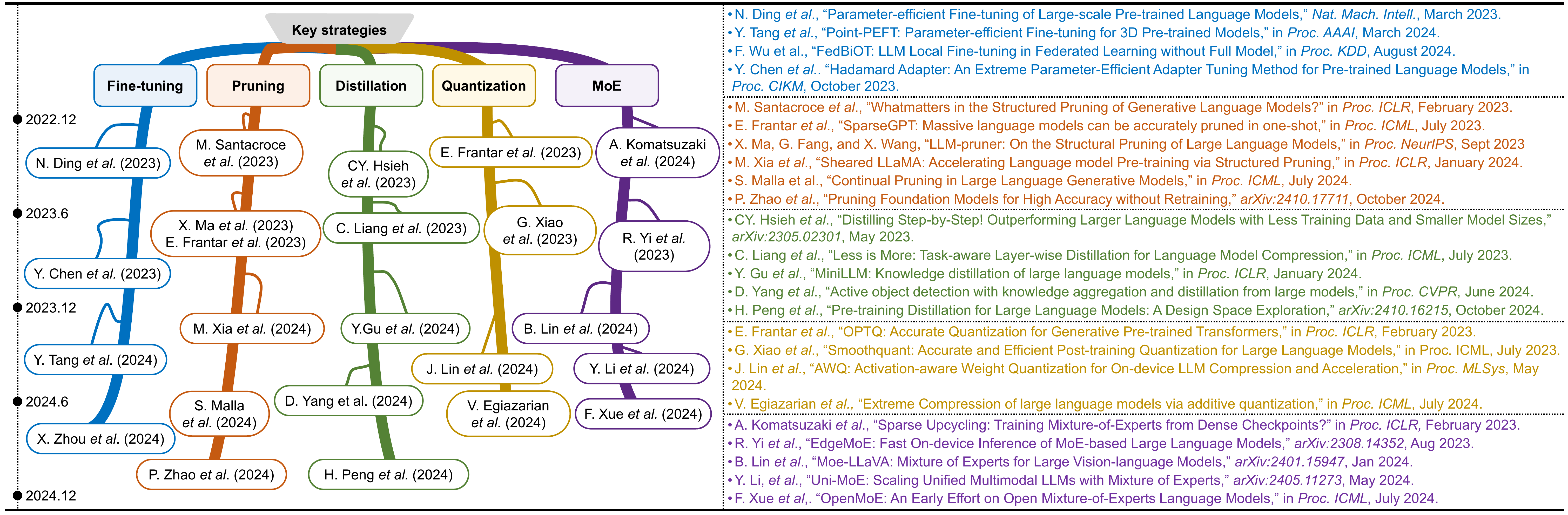}}
\caption{Review of representative research works on key compact model in democratized generative AI. From the tree diagram, we can find that in the past two years, many research works have emerged to improve and optimize large generative AI models such as LLMs, which will further promote the progress and realization of the democratized generative AI paradigm in the next-generation mobile edge networks.}\label{Fig2}
\end{figure*}

To study the research trends of the above five key compact model strategies in democratized generative AI, we also present a tree diagram that reviews some representative research works in the recent two years (i.e., 2022 to 2024), as shown in Fig. \ref{Fig2}. Note that a comprehensive survey and exhaustive list of works can be future work.

\subsection{Extended Network Applications under Compact Model Strategy Coordination}
An efficient combination of different compact model strategies and network architectures can promote the development of next-generation networking. Therefore, we discuss three major network applications to illustrate their synergistic effects.

\subsubsection{Cloud-Edge Collaborative Networks}
Compact model strategies, such as pruning, quantization, and distillation, are essential for enabling cloud-edge collaborative networks, which are crucial for running resource-intensive generative AI models on resource-constrained mobile edge devices. Pruning and quantization effectively reduce model complexity, allowing mobile edge devices to handle simpler tasks locally, while distillation enhances model efficiency to support cloud-edge integration. Together, these strategies optimize resource allocation, enabling mobile devices to participate actively in cloud-edge collaboration without excessive dependency on the cloud. For example, the GKT\footnote{\url{https://github.com/Zoeyyao27/GKT}} framework employs knowledge distillation to assign larger teacher models to cloud servers and deploy smaller student models on mobile edge devices, allowing them to work collaboratively. This can reduce data transmission and support efficient cloud-edge collaboration in LLM deployment.

\subsubsection{Adaptive Decision-Making for Multi-Agent Systems}
Compact model strategies, particularly knowledge distillation and Mixture of Experts (MoE), are instrumental in enabling adaptive decision-making for multi-agent network systems. These strategies facilitate decentralized decision-making and effective communication across distributed networks, allowing agents to leverage distilled knowledge without requiring extensive data exchange. This contributes to building scalable and resilient multi-agent environments.  For example, in \cite{zou2024genainet}, the authors presented GenAINet, a semantic-native framework for wireless device queries that uses knowledge distillation to transfer knowledge among multiple generative AI agents, enabling collaborative on-device query processing under the device-to-device communication paradigm.

\subsubsection{Optimization in Vertical Network Fields}
Compact model strategies, such as fine-tuning and MoE, can enable LLMs to specialize for specific vertical fields within networking like network design, diagnosis, and configuration~\cite{huang2024large}. In this context, fine-tuning allows LLMs to adapt to network-specific terminology, enabling them to interpret technical instructions accurately, respond to network events, or assist in troubleshooting. Likewise, MoE structures selectively activate domain-relevant knowledge bases or sub-models, dynamically adjusting LLMs' responses based on context. These strategies together facilitate applications such as network automation, adaptive configuration, and tool integration, allowing generative AI to support real-time decision-making in complex network environments. For instance, Mobile-LLaMA\footnote{\url{https://github.com/DNLab2024/Mobile-LLaMA}} is an LLM fine-tuned on a network-specific dataset, supporting functions such as packet analysis and IP routing analysis, thereby providing practical tools for generating code scripts tailored to 5G network analysis.

\begin{figure*}[t]
\centering{\includegraphics[width=0.88\textwidth]{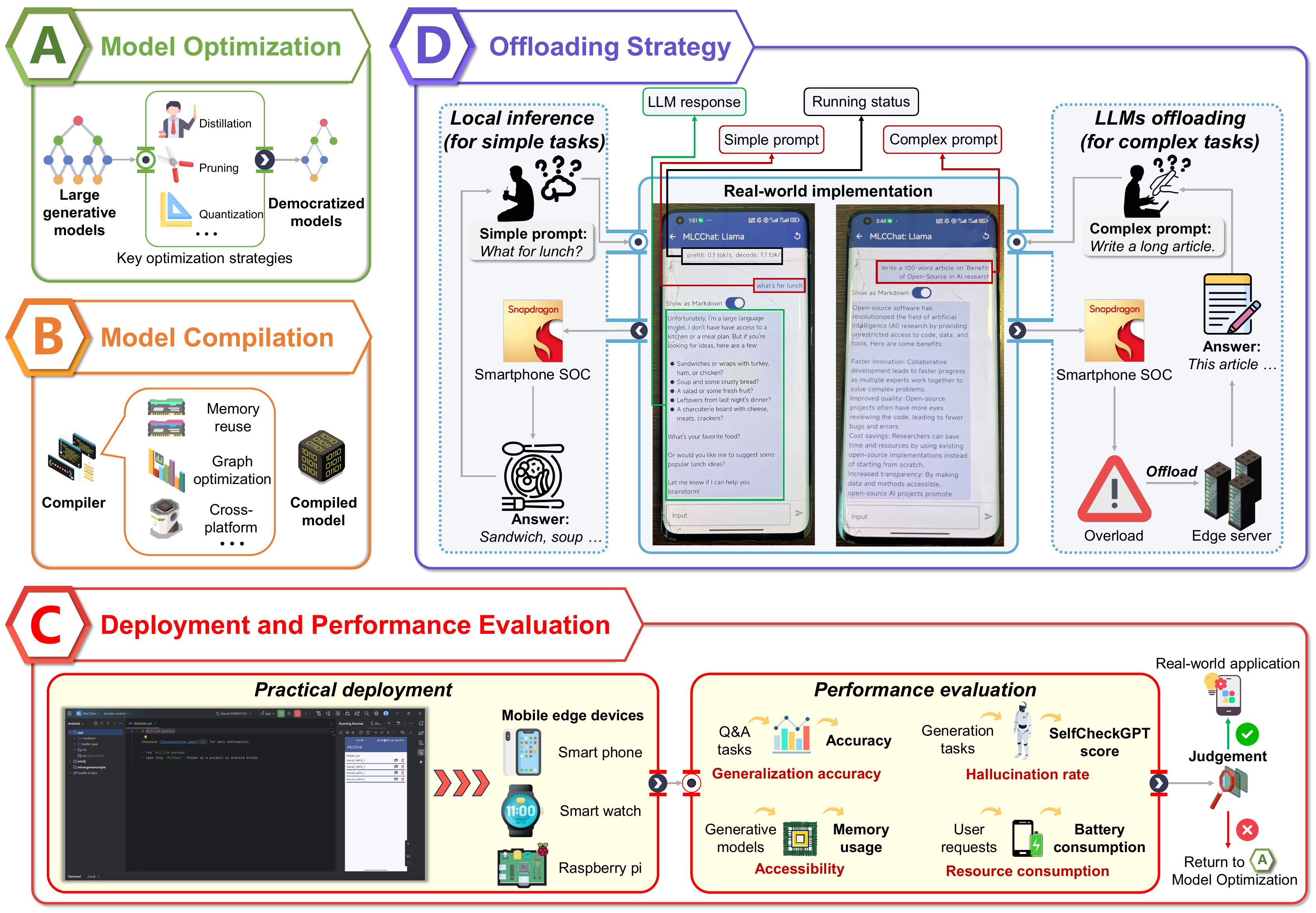}}
\caption{Workflow for deploying LLMs on mobile edge devices for real-time Q\&A tasks. Key steps include (A) \textbf{Model Optimization} using techniques such as pruning and quantization; (B) \textbf{Model Compilation} to optimize models for hardware compatibility; (C) \textbf{Deployment and Performance Evaluation} with metrics such as accuracy, hallucination rate, and resource usage; (D) \textbf{Offloading Strategy}, where simple tasks are handled locally and complex tasks are offloaded to an edge server, balancing efficiency and scalability.}
\label{Fig3}
\end{figure*}

\section{Case Study: Democratized LLMs for Real-Time Decision-Making on Mobile Edge Devices}

As illustrated in Fig.~\ref{Fig3}, we present a case study for realizing democratized generative AI through LLM deployment on mobile edge devices. Furthermore, we conduct experiments to evaluate the effectiveness of the edge LLM deployment.

\subsection{Tutorial Workflow}
The workflow for deploying LLMs on mobile edge devices involves the following four steps:

 \textbf{Step 1: Model Optimization.}  
    The first step is optimizing the model to run efficiently on resource-constrained devices. This process includes downsizing the LLMs using compact strategies such as quantization, model pruning, and knowledge distillation. In this case study, we apply quantization, distillation, pruning, and fine-tuning separately to Llama models, resulting in seven distinct candidate models, each with unique advantages. For instance, the 4-bit quantized Llama-3.1-8B model compresses its parameter size to a quarter of the original, greatly reducing the memory needed to store and transport its parameters. Meanwhile, fine-tuned models enhance generalization accuracy and reduce hallucination rates.

\textbf{Step 2: Model Compilation.} 
    After optimizing the model, we compile it to further optimize its performance for specific hardware of mobile edge devices. In this case study, we use compilation methods provided by MLC-LLM\footnote{\url{https://llm.mlc.ai/}} to generate binary model libraries, lightweight runtime, and tokenizers. By utilizing the tensor virtual machine (TVM), the model is transformed into a more hardware-optimized form, which reduces latency and energy consumption during inference.  Finally, we use Android Studio to build the \textit{.apk} file to make it compatible with the Android platform. The compiled models are specifically adjusted to fully utilize the processing capabilities of target hardware, such as mobile CPUs and GPUs.

 \textbf{Step 3: Deployment and Performance Evaluation.}
    After compilation, the model is deployed to mobile edge devices and subjected to a comprehensive evaluation to ensure that it meets the necessary performance metrics. This step involves assessing metrics, including generalization accuracy, hallucination rate, resource consumption, and accessibility. In this case study, we test the hallucination rate and generalization accuracy of these models using text generation tasks and Q\&A tasks, respectively. The accessibility and resource consumption is evaluated by the memory footprint of the model parameters and the energy consumption caused by model inference, respectively. Models that fail to meet the requirements are iteratively returned to \textbf{Step 1} for further adjustments and optimizations. Only after passing these evaluations is the model deployed in a real-world scenario.

  \textbf{Step 4: Offloading Strategy.}  
    To balance performance and resource consumption, we implement an offloading strategy where unquantized versions of the LLMs are deployed on an edge server. This approach allows lightweight models on mobile devices to handle basic requests locally, ensuring fast responses and low latency. For more complex tasks that exceed the capabilities of local models, the requests are offloaded to the edge server, where full-scale LLMs process them. This hierarchical approach ensures optimal efficiency and scalability, providing a better user experience while maximizing resource utilization.


\textbf{Simulation Settings}: To assess the performance of the democratized generative AI, we evaluate the model's generalization accuracy, hallucination rate, accessibility, and resource consumption as described in {\textbf{Step 3}}. Specifically, we perform Q\&A tasks using LLMs to evaluate their generalization accuracy. The \textit{WebQuestionsSP} dataset, which contains 4,737 questions\footnote{\url{https://www.microsoft.com/en-us/download/details.aspx?id=52763}}, is used, and each question is asked independently. A model is considered to answer correctly if its response includes the correct answer. 
To evaluate the hallucination rate, we use LLMs to generate content similar to that in the Wiki Bio GPT-3 Hallucination dataset\footnote{\url{https://huggingface.co/datasets/potsawee/wiki_bio_gpt3_hallucination}} and calculate the hallucination rate by SelfCheckGPT with NLI score~\cite{manakul2023selfcheckgpt}. The hallucination rate is expressed as a value between 0 and 1, with lower values indicating better performance. The accessibility of democratized generative AI models in edge networks is mainly constrained by the available memory of edge devices. Therefore, we assess the memory requirements of the LLMs, where the lower memory requirements indicate higher accessibility. Additionally, we calculate the average battery consumption of the mobile edge devices when LLMs respond to a single query to evaluate their resource consumption.

\subsection{Numerical Results}
\begin{figure}[t]
\centering{\includegraphics[width=0.48\textwidth]{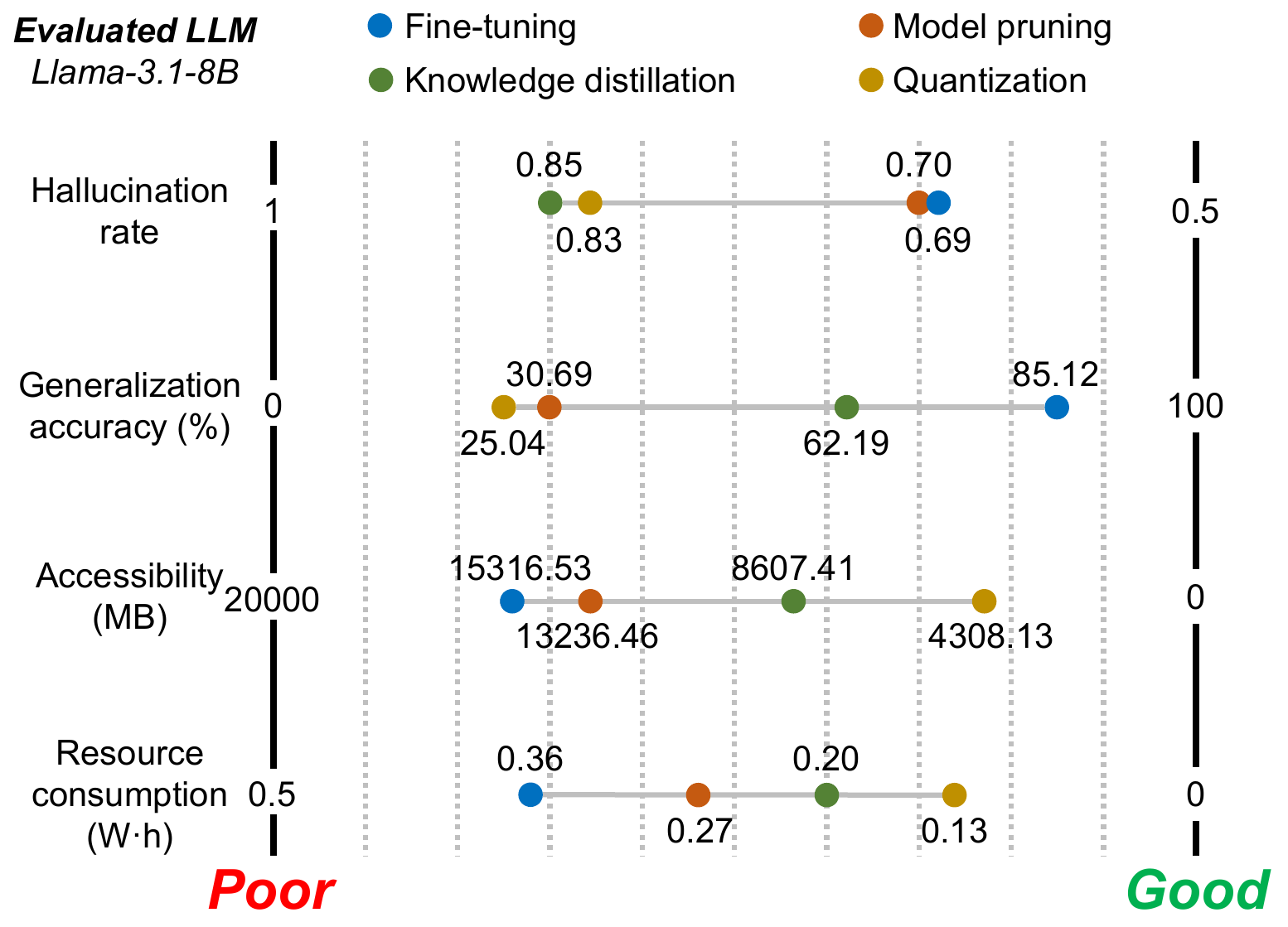}}
\caption{Experimental results of key compact model strategies on the metrics. This test is conducted on a server with Intel Xeon Platinum 8380 CPU and NVIDIA A100 80GB GPU. Resource consumption is the average power consumption of the GPU to respond to a single query with up to 256 tokens.}\label{Fig4}
\end{figure}

For comparison, we conduct experiments on the Llama-3.1-8B model using four compact generative AI strategies: \textit{fine-tuning}\footnote{\url{https://huggingface.co/meta-llama/Llama-3.1-8B-Instruct}}, \textit{model pruning}\footnote{\url{https://huggingface.co/Na0s/Llama-3.1-8B-Pruned-4-Layers}}, \textit{4-bit quantization}\footnote{\url{https://huggingface.co/mlc-ai/Llama-3.1-8B-q4f16_1-MLC}}, and \textit{knowledge distillation}\footnote{\url{https://huggingface.co/nvidia/Llama-3.1-Minitron-4B-Width-Base}}. As shown in Fig.~\ref{Fig4}, the hallucination rate demonstrates a negative correlation with parameter size (i.e., accessibility), where the larger models tend to exhibit lower hallucination rates. Furthermore, the models optimized using distillation and fine-tuning outperform those using quantization and pruning in terms of generalization accuracy. This is because distillation and fine-tuning involve a training process where developers can guide the model through loss functions to improve correctness. While fine-tuning achieves high generalization accuracy, it does not reduce the parameter size, resulting in poorer accessibility and higher energy consumption. On the other hand, quantization offers the highest accessibility and resource efficiency, though it results in a higher hallucination rate and lower generalization accuracy compared to other methods. Pruning, similar to quantization, achieves better accessibility and energy efficiency by reducing the parameter size, but it incurs a greater loss in generalization accuracy than those of the other strategies.

\begin{figure*}[t]
\centering{\includegraphics[width=0.98\textwidth]{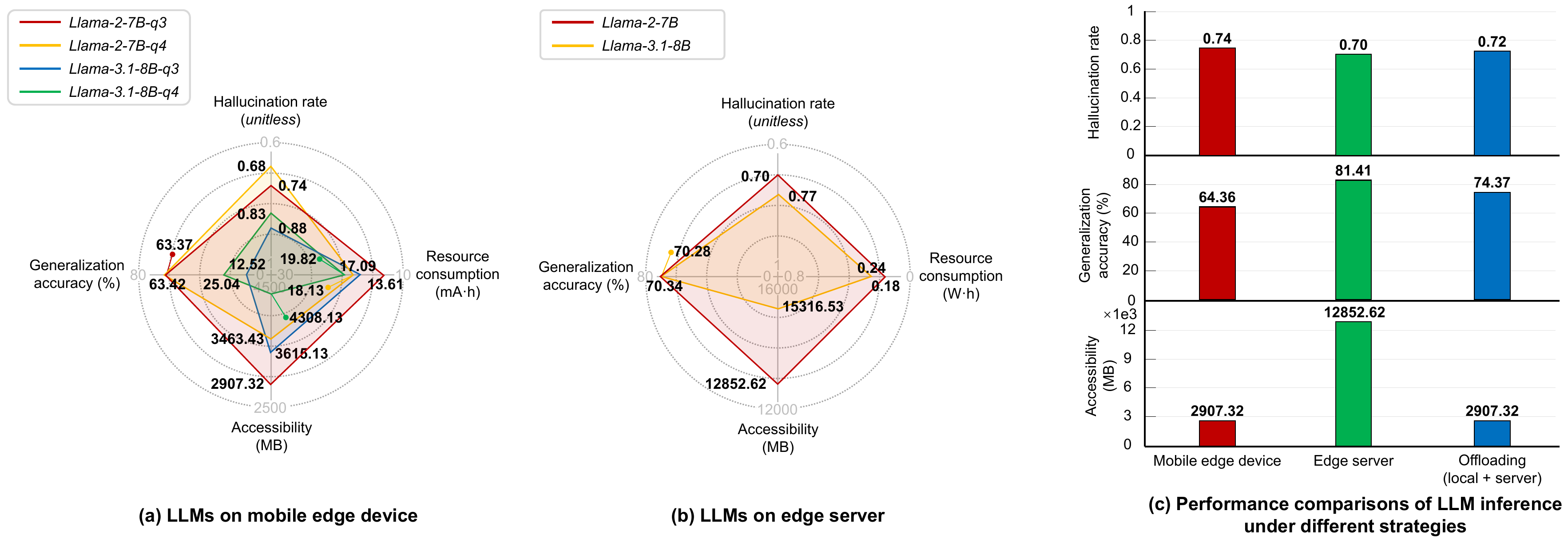}}
\caption{Numerical results of running the democratized LLMs. The mobile edge device used is a Xiaomi 10 Ultra smartphone with 12GB of RAM, 256GB of storage, and a Qualcomm Snapdragon 865 processor. The edge server is with Intel Xeon Platinum 8380 CPU and NVIDIA A100 80GB GPU. The Resource consumption of a mobile edge device is the average battery consumption to respond to a single query, while the edge server resource consumption is the average power consumption of the GPU to respond to a single query with up to 256 tokens.}\label{Fig5}
\end{figure*}

Next, we explore the impact of different parameter settings of compact strategies on the key metrics. As shown in Fig.~\ref{Fig5}(a), we take quantization as an example strategy to examine how different parameter settings within the same compact model strategy affect model performance. It is observed that while increasing the degree of quantization results in improved resource consumption and accessibility, it also leads to a higher hallucination rate and decreased generalization accuracy. Specifically, the models with 3-bit quantization achieve better accessibility and lower resource consumption than those with 4-bit quantization but at the cost of increased hallucination rates. Thus, selecting an appropriate optimization setting that balances accessibility and performance is critical, especially given the resource constraints of mobile edge devices. Furthermore, we evaluate the performance of LLMs on an edge server on performance metrics. As shown in Fig.~\ref{Fig5}(b), the Llama-3.1-8B model performs equally well as Llama-2-7B-chat in both hallucination rate and generalization accuracy. However, in Fig.~\ref{Fig5}(a), the performance of the quantized Llama3.1-8B model has significantly decreased. In particular, the 3-bit quantized Llama-3.1-8B model reaches only 12.52\% accuracy on the WebQuestionsSP dataset. This highlights the importance of comprehensive performance evaluation before deployment, such a model should be returned to \textbf{Step 1} for further optimization to meet the requirements for deployment in resource-constrained environments.

Finally, following \textbf{Step 4}, we set up an offloading scenario using 3-bit quantized and original Llama-2-7B-chat models deployed on mobile edge devices and edge servers, respectively. We use single-hop and multi-hop questions from the WebQuestionsSP dataset to serve as simple and complex prompts. In the offloading setup, simple questions are processed locally by the mobile edge device, while complex questions are offloaded to the edge server for handling. We also compare the hallucination rate, generalization accuracy, and accessibility of three strategies: processing entirely on the mobile edge device, processing entirely on the edge server, and using the offloading strategy (i.e., local device + edge server). As shown in Fig.~\ref{Fig5}(c), the accessibility of the offloading strategy remains identical to that of running entirely on the mobile edge device, as both scenarios require the same memory footprint for loading models on mobile devices. However, the offloading strategy shows an improved hallucination rate and generalization accuracy compared to the purely local strategy, as complex tasks are offloaded to the edge server, where more powerful models are available.



\section{Future Directions}


\textbf{Security and Privacy Challenges:}
Ensuring both the security and privacy of data during generative AI models or LLMs offloading remains a critical challenge in mobile edge environments. Sensitive data transmitted between mobile devices and edge servers is at risk of tampering and unauthorized access, which could lead to compromised model responses or leakage of sensitive user information. To address these issues, robust encryption techniques and secure communication protocols are necessary to protect data integrity and confidentiality during transmission. Additionally, privacy-preserving methods, such as differential privacy and federated learning, should be explored to minimize the risk of eavesdropping and data leakage, building trust and maintaining user privacy in mobile-edge AI applications.

\textbf{Networking Challenges:}
Ensuring efficient and reliable network communication is crucial when deploying generative AI models or LLMs in mobile and edge environments, where network instability can impact performance. Limited bandwidth, latency variability, and potential congestion in mobile networks can lead to delayed responses or reduced model effectiveness, especially during high-demand scenarios. Addressing these challenges calls for developing adaptive networking protocols that optimize data transfer, reduce latency, and support seamless offloading under dynamic network conditions. Future research should focus on intelligent traffic management, predictive data caching, and network resource allocation to maintain smooth AI services across varying network environments.

\textbf{Accuracy Enhancement:}
Accuracy is a key concern when deploying downsized LLMs or generative AI models on mobile devices, as reducing the model size often results in degraded performance. Addressing this issue requires the development of advanced optimization techniques that can enhance the accuracy of lightweight models while retaining their reduced footprint. Future research should explore adaptive learning strategies to maintain high accuracy in resource-constrained settings. 

\section{Conclusion}
In this article, we have explored the democratization of generative AI deployment on mobile and edge networks. We have reviewed compact strategies for democratizing AI, such as quantization, model pruning, and knowledge distillation, focusing on their effectiveness in optimizing LLMs for resource-constrained environments. We have conducted a case study deploying LLMs on real mobile edge devices, demonstrating the practicality of democratized AI through significant improvements in generalization accuracy, hallucination rate, accessibility, and resource consumption. Finally, we have outlined potential research directions to further advance the democratization of generative AI for mobile and edge scenarios.

\bibliographystyle{IEEEtran}
\bibliography{mylib}

\end{CJK}
\end{document}